\documentclass[pra,a4paper,twocolumn,showpacs,preprintnumbers,amsmath,amssymb,floatfix,amstex,superscriptaddress]{revtex4}
\newcommand{\ket}[1]{|#1\rangle}                	%
\newcommand{\brac}[1]{\langle #1|}              	%
\newcommand{\hil}{{\cal H}}              		%
\newcommand{\hilNN}{{\cal H}_{N^2}}          		%
\newcommand{\De}{\text{\textbf{\textsf{D}}}_{\sigma}} 	
\newcommand{\Tqp}{\hat{T}_{(q,p)}}           		%
\newcommand{\hrho}{\hat{\rho}}               		%
\newcommand{\hA}{\hat{A}}                		%
\newcommand{\hB}{\hat{B}}                		%
\newcommand{\hM}{\hat{M}}                		%
\newcommand{\ca}{C_{\sigma}(\alpha)}             	%
\newcommand{\cl}{\widetilde{C_{\sigma}}(\lambda)}  	%
\newcommand{\cmn}{\widetilde{C_{\sigma}}(\mu,\nu)}      %
\newcommand{\SOp}[1]{\text{\textbf{\textsf{#1}}}}       
\newcommand{\doll}{\text{\textbf{\textsf{L}}}} 		
\newcommand{\II}{\hat{I}}        			%
\newcommand{\s}{\text{\textbf{\textsf{S}}}}		%
\newcommand{\hT}{\hat{T}}               		%
\newcommand{\hU}{\hat{U}}       			%
\newcommand{\hV}{\hat{V}}       			%
\newcommand{\hW}{\hat{W}}       			%
\newcommand{\hg}{\hat{\gamma}}       			%
\newcommand{\hQ}{\hat{Q}}       			%
\newcommand{\hUd}{\hU^\dagger}              		%
\newcommand{\tr}{\mbox{Tr}}                 		%
\newcommand{\equa}[1]{Eq.~(\ref{#1})}       		%
\newcommand{\sect}[1]{Sec.~\ref{#1}}			%
\usepackage{graphicx,graphics,wrapfig,rotating}     	
\usepackage{dcolumn}                    		
\usepackage{bm,fancybox}                		
\usepackage{times,euscript,eufrak,oldgerm}              %
\usepackage[german,english]{babel}			%
\begin{document}
\title{Noise models for superoperators in the chord representation} %
\author{Mario Leandro Aolita}
\altaffiliation[Present address: ]{%
Instituto de F\'\i sica, Universidade Federal do Rio de Janeiro. Caixa Postal 
68528, 21941-972 Rio de Janeiro, RJ, Brasil.\\
Email address: \url{aolita@if.ufrj.br}}
\affiliation{%
Departamento de F\'{\i}sica, Comisi\'{o}n Nacional de Energ\'{\i}a At\'{o}mica.
Avenida del Libertador 8250 (C1429BNP), Buenos Aires, Argentina.
}
\author{Ignacio Garc\'\i a-Mata}
\email[Email address: ]{garciama@tandar.cnea.gov.ar}
\affiliation{%
Departamento de F\'{\i}sica, Comisi\'{o}n Nacional de Energ\'{\i}a At\'{o}mica.
Avenida del Libertador 8250 (C1429BNP), Buenos Aires, Argentina.
}%
\author{Marcos Saraceno}
\email[Email address: ]{saraceno@tandar.cnea.gov.ar}
\affiliation{%
Departamento de F\'{\i}sica, Comisi\'{o}n Nacional de Energ\'{\i}a At\'{o}mica.
Avenida del Libertador 8250 (C1429BNP), Buenos Aires, Argentina.
}%
\affiliation{%
Escuela de Ciencia y Tecnolog\'\i a, Universidad Nacional de San
Mart\'\i n. Alem 3901 (B1653HIM), Villa Ballester, 
Argentina.}
\date{\today}%
\begin{abstract}
We study many-qubit generalizations of quantum noise channels that can be written
as an incoherent sum of translations in phase space, for which the chord representation
results specially useful. 
Physical description in terms of the spectral properties of the superoperator 
and the action in phase space are provided. 
A very natural description of decoherence leading to a preferred basis is achieved with
diffusion along a phase space line. The numerical advantages of using the chord
representation are illustrated in the case of coarse-graining noise.
\end{abstract}
\maketitle

\section{Introduction}
The precise manipulation of coherent quantum processes is the ultimate goal at the basis of quantum
information theory. However experimental quantum systems cannot be completely isolated and therefore
unwanted interactions pose strict limits on its practical factibility.

Therefore 
understanding the effects of certain kinds of noise on 
algorithms and how to correct the
errors they produce is a major subject of interest.

The purpose of this paper is to show how some simple noise models of interest in quantum
information theory can be formulated in a phase space language. In doing so we reinterpret
the depolarizing and the phase damping channels on many qubits as a special kind of
diffusion in phase space, study their spectral properties and display their effects on
selected pure initial states.

The Weyl
representation, and its associated Wigner function, have been successfully
used to understand many aspects of quantum mechanics \cite{davi}, 
especially its classical limit \cite{zurek1,paz1}. In its discrete version,
it has also been used to study the properties of quantum maps and, recently,
to analyze {\em many-qubit} quantum algorithms \cite{Miquel}. In phase space the Wigner
distribution of a state displays both its classical properties, in the
smooth part of the distribution, as well as its quantum properties, in the form of
highly correlated oscillatory structures with sub-Planck scale sizes
\cite{zurek_nature}. To reveal these structures ocurring at widely different scales
it is then natural to consider the Fourier transform of the Wigner function. This
procedure leads to a different, and alternative, representation of quantum
mechanics in phase space. This representation, when applied to density matices is
generally called {\em  characteristic\/} or {\em generating function\/} representation.
Its properties as a representation for general operators have been extensively
studied in \cite{ozorivas} where due to its geometrical features it has been
called the {\em chord representation\/}. We adhere to this nomenclature. In
this paper we show the advantages and the simplicity of using this representation
as a ``noise'' basis, where various models of noise are given a simple phase space
interpretation.

The paper is organized as follows.
In \sect{sec:2stage} we briefly review some fundamental concepts about quantum
open systems and propose a two-stage scheme for the noisy propagator 
of density operators. In \sect{sec:chordrep} we describe noise models in phase
space. Particularly, noise superoperators which are diagonal in the chord or
characteristic representation \cite{ozorivas} or, as we show,
whose Kraus operators are proportional to translation operators in phase space. 
In \sect{sec:channels} we find generalizations of the depolarizing and phase
damping channels \cite{preskill,chuang} for many qubits using the chord
representation. Finally, in \sect{sec:difnoise} we describe the diffusive noise recently
proposed in
\cite{garma,garma2} and show how it can be used to obtain the main part of the
spectrum of the noisy propagator. 
\section{A model of noisy evolution}
\subsection{Two-stage Superoperators}
		\label{sec:2stage}
Quantum systems coupled to a Markovian environment evolve according to the
master equation \cite{L,qnoise}
\begin{equation}
	\label{eq:master}
\frac{d\hat\rho}{d t}=-{i\over\hbar}[\hat
H,\hrho]+{1\over\hbar}\sum_j\left(\hat
L_j\hrho\hat L_j^\dag -\frac{1}{2}\hat L_j^\dag\hat L_j\hrho
-\frac{1}{2}\hrho\hat L_j^\dag\hat L_j\right), 
\end{equation}
where $\hat L_j$ are the Lindblad operators. The first term on the right gives
the unitary evolution. The second term represents interaction with an
environment. Equation
(\ref{eq:master}) generates a solution in the form of a one parameter family
({\em quantum dynamical semigroup\/} \cite{GKS})
of linear maps  
or superoperators $\doll_t$,
such that the state at time $t$ is given by
\begin{equation}
\hrho_t=\doll_t(\hrho_0).
\end{equation}
A general way of representing a superoperator is as a
{\em Kraus operator sum\/} \cite{Kraus}. Given a superoperator
$\doll$ there exist a set of (Kraus) operators $\hat{M}_\mu$,
such that
\begin{equation}
\doll=\sum_{\mu}\hM_{\mu}\odot\hM^{\dag}_{\mu},
\end{equation}
where, for $\doll$ to be trace preserving, the identity
$\sum_{\mu} \hM_{\mu}^\dag\hM_{\mu}=\II$ must hold. 
Throughout the paper the
$\odot$ symbol should be interpreted as
\begin{equation}
\hM\odot\hM^\dag(\hrho)\stackrel{{\rm
def}}{=}\hM\hrho\hM^\dag. 
\end{equation}
Not only does the Kraus form ensure
that $\doll(\hrho)$ is positive, for any density
operator $\hrho$, but it is also {\em completely positive\/},
meaning that tracing over any environment on which $\doll$
acts trivially yields again a density operator \cite{havel}. 

In this paper instead of modeling environments and solving the master equation
we focus on the properties of  $\doll$ for open quantum 
systems and thus propose models for noisy propagators. This shift of emphasis is analogous
to the shift from hamiltonians to quantum maps in the study of unitary dynamics.

We center our attention on discrete time systems 
\begin{equation}
\hrho_n=\doll(\hrho_{n-1})=\doll^n(\hrho_0)
\end{equation} 
and propose a two-stage propagator consisting on the composition of a unitary
with a noisy evolution.
After one unitary step a density matrix $\hrho$ becomes
\begin{equation}
\hU\hrho\hU^\dag\stackrel{\rm def}{=}\SOp{U}(\hrho). 
\end{equation}
The operator $\SOp{U}$ is a unitary superoperator
with trivial Kraus form. 
The interaction with a Markovian environment is modeled by a noise superoperator
$\De$ depending on a parameter that quantifies its strength.
 The noisy one-step propagator is thus defined as
\begin{equation}
		\label{eq:noisyprop}
\doll(\hrho)=\De\circ\SOp{U}(\hrho).
\end{equation}
Similar types of two-stage schemes can be found in
\cite{voros,braun,nonn,garma,garma2,spina}. 
There are several situations where this scheme appears naturally. One example is a
kicked map in which the noise only acts between kicks. A billard inside 
a bath  constitutes another example if we consider
that the interaction with the walls is purely unitary while the free propagation
is noisy. A quantum algorithm, supposed perfect, sent through a noisy channel is another example.
 In the case of a quantum algorithm that needs to be iterated, like Grover's search,
the noisy part would be an effective interaction acting after each iteration. Finally a
 numerical
solution of Lindblad's equation proceeding in small time steps will naturally alternate
between the unitary and the noisy propagation.

\subsection{Diagonal noise in the chord representation}
		\label{sec:chordrep}
Systems of $K$ qubits are usually treated in a tensor product basis, 
the computational basis,
defined as the tensor product of the eigenstates of $\hat\sigma_z$,
$\{\ket{1},\ket{0}\}$. It
is labeled by an integer $\ket{n}$ ranging from $n=0,...N-1$ where $N=2^K$
(throughout the paper, for many qubit systems, we call $K$ the number of qubits
and $N$ the dimension of phase space). 
The coeficients of the binary expansion of $n$ represents the state of each single qubit. 
As an operator basis,
it is also customary to take the tensor product	of Pauli operators, consisting  of the 
$N^2=4^K$ possible tensor products of $ 1, \sigma_x,\sigma_y, \sigma_z $.
	
An alternative approach, standard in the treatment of quantum maps \cite{ozorivas}, 
and also
recently utilized in the context of quantum information \cite{nature,Miquel,Paz} 
is to treat $K$-qubit systems using a phase space setting. 
In this approach the computational basis is assimilated to the position basis
$B_q=\{\ket{n}, n=0,\cdots,N-1\}$ 
(with periodic boundary conditions). 
Therefore the conjugate discrete momentum basis
$B_p=\{\ket{k},k=0,\cdots,N-1\}$
is obtained by means of the discrete Fourier transform (DFT). 
The dimension $N$ is then naturally related to an effective Planck constant by
\begin{equation}
N=1/2\pi\hbar.
\end{equation}

The natural operator basis in this context is constituted by 
the phase space translations
\cite{schwinger,ozorivas,Miquel} $\Tqp$ defined as
\begin{eqnarray}
\Tqp\ket{k}&=&\exp\left[-(2\pi i/N)
q\left(k+p/2\right)\right]\ket{k+p}\\
		\label{eq:tqp}
\Tqp\ket{n}&=&\exp\left[(2\pi i/N)
p\left(n+q/2\right)\right]\ket{n+q},
\end{eqnarray}
with the group composition rule
\begin{equation}
    \label{eq:group}
\hat{T}_{(q_1,p_1)}\hat{T}_{(q_2,p_2)}=\hat{T}_{(q_1+q_2,p_1+p_2)}\,
e^{
(i\pi/N)(p_1 q_2-q_1 p_2)},
\end{equation}
where the phase in the exponential is the area of the triangle
defined by the vertices $(0,0)$, $(q_1,p_1)$ and $(q_2,p_2)$. 
Translations can be written in terms of compositions of translations in position and momentum. 
From \equa{eq:group},
\begin{eqnarray}
	\label{eq:Tdecomp}
\Tqp&=&U^qV^p\,e^{i(\pi/N)q p}\nonumber \\
    &=&V^pU^q\,e^{-i(\pi/N)q p}
\end{eqnarray}
where
\begin{equation}
  \begin{array}{rll}
  U^q&\stackrel{\text{def}}{=}&\hT_{(q,0)},\\
  V^p&\stackrel{\text{def}}{=}&\hT_{(0,p)}.
  \end{array}
\end{equation}
For the sake of simplicity,
from now on we use a single (greek letter) index to
represent the $N^2$ $(q,p)$ points of the discrete phase space, except where two
indeces are needed explicitly. From
\equa{eq:group} it is easy to show that 
\begin{equation}
\hT_{\alpha}^\dag\hT^{ }_{\alpha}=\II.
\end{equation}
These operators have matrix representation in
$\hilNN$ or ``Liouville'' space, the space of  $N\times N$ complex matrices 
with the Hilbert-Schmidt inner product
\begin{equation}
	\label{eq:inner}
(\hat{A},\hat{B})={\rm Tr}(\hat{A}^\dag\hat{B}),
\end{equation}
where $\hA$ and $\hB\in\hilNN$ (we use ``matrix'' or 
``operator'' to refer to elements in $\hilNN$ indistinctly, 
except where an explicit distinction is required). 
There are $N^2$ operators $\Tqp$ in $\hilNN$ and 
they form a complete orthogonal set since
\begin{equation}
\tr(\hT_\alpha^\dag\hat{T}_{\alpha'}^{ })=N\delta_{\alpha\alpha'}.
\end{equation}
Therefore
any operator $\hA$ in $\hilNN$ can be expanded in this unitary basis as
\begin{equation}
    \label{eq:chord}
 \hA=\frac{1}{\sqrt{N}}\sum_{\alpha}a(\alpha)\hT_{\alpha}.
\end{equation}
The $c$-number function (or symbol)
\begin{equation}
a(\alpha)=\frac{1}{\sqrt{N}}\tr (\hA \hT_{\alpha}^\dag)
\end{equation}
defines the {\em chord\/} representation \cite{ozorivas} of $\hA$.
The DFT of the translation operators (with the dimension extended to $2N$,
see \cite{hannay}) are the generalized phase space
point operators. Operators written in this basis constitue the Weyl or {\em center\/}
\cite{ozorivas} representation.
The symbol of a density operator in the center representation is the well known discrete
Wigner function, while in the chord representation it is also known as generating or
characteristic function. 
A description of the phase space point operators and the peculiar features
of the discrete Wigner function can be found in \cite{Miquel}.

A general superoperator $\SOp{D}$ can be written in the basis of translations as 
\begin{equation}
	\label{eq:Texpansion}
\SOp{D}=\frac{1}{N}\sum_{\alpha,\beta}C(\alpha,\beta)\hT^{ }_\alpha\odot\hT^\dag_\beta.
\end{equation} 
where the only requirement is that the matrix coeficients  
$C(\alpha,\beta)$ be non-negative and $C$ has unit trace.
In this general setting the Kraus operators are linear superpositions of translations
obtained by diagonalizing  $ C(\alpha,\beta)$. In this paper we only consider 
superoperators where  $ C(\alpha,\beta)=\delta_{\alpha,\beta}C(\alpha)$ is already in
diagonal form in such a way that the Kraus operators are simply proportional to the
translations  $ \hT^{ }_\alpha $.
Thus, the noise superoperator that we consider is explicitly 
defined by giving the Kraus operator sum form
\begin{equation}
    \label{eq:deps}
\De=\frac{1}{N}\sum_\alpha C_\sigma(\alpha) 
\hT_\alpha^{ }\odot\hT_\alpha^\dag.
\end{equation}
Trace preservation is achieved if 
\begin{equation}
\sum_{\alpha}\ca/N=1.
\end{equation}
From \equa{eq:deps} we see that $\De$
is a convex sum of unitaries yielding a contracting superoperator.
Physically the action of $\De$
can be simply interpreted as  performing an incoherent sum of all the possible translations
$\hT_\alpha$ each with a probability $\ca/N$. In this way the choice of $\ca$
determines different types of noise. The parameter $\sigma$ is introduced to control
the {\em strength\/} of the noise.

One can think of this as a noise channel where the ``errors'' are unitary translations in
phase space where they occur with  probability $\ca/N$ . In this sense it constitutes a
``very nice'' \cite{knill} error basis with properties that are different from the more usual 
ones given by tensor products of Pauli matrices.

One of the motivations for using noise in the form of \equa{eq:deps} is
that all the spectral properties are readily available. From \equa{eq:group} it is
clear that  
\begin{eqnarray} 
\label{eq:spectDe}
\De(\hT_\lambda)&=&\frac{1}{N}\sum_\alpha C_\sigma(\alpha)
	\hT_\alpha\hT_\lambda\hT_\alpha^\dag\nonumber \\ 
		&=&\frac{1}{N}\sum_\alpha
	C_\sigma(\alpha) e^{-i(2\pi/N)\lambda\wedge\alpha}
	\hT_\lambda \nonumber \\
		&=&\cl \hT_\lambda 
\end{eqnarray} 
where $\wedge$ is the {\em wedge product\/} $\lambda\wedge\mu=\mu p-\nu q$, with 
$\lambda=(\mu,\nu)$ and
$\alpha=(q,p)$. Then the 
eigenfunctions of $\De$ are the translation operators $\hat{T}_{\lambda}$
and the corresponding eigenvalues are given by $\cl$,  the DFT of $\ca$.
As $\De$ is diagonal
in the chord representation (\ref{eq:chord}) its action is quite
simple. If $\hrho$ is expanded as 
\begin{equation}
\hrho=\frac{1}{\sqrt{N}}\sum_{\lambda}\rho_{\lambda}\hat{T}_{\lambda}
\end{equation} 
then 
\begin{equation}
\De(\hrho)=\frac{1}{\sqrt{N}}
\sum_{\lambda}\cl\rho_{\lambda}\hat{T}_{\lambda}.
\end{equation} 
In the chord representation the action of $\De$ is simply to 
{\em modulate\/} the elements $\hrho_{\lambda}$ with $\cl$.

It is also of considerable interest to determine the spectral properties of the
combined action of $\De$ with a unitary step $\SOp{U}$ as in
\equa{eq:noisyprop}. In many instances  (see sections \ref{sec:channels} and
\ref{sec:difnoise})
a significant portion of the noise spectrum $\cl$ is zero or is contained within
a small boundary of zero in the complex plane, with just a few isolated
eigenvalues in the annular region between it and the unit circle. This
quasi-null subspace reduces the effective rank of the combined supeoperator
$\De\circ\SOp{U}$ and allows a very efficient calculation of the leading
spectrum with diagonalizations of relatively small size (see
\sect{sec:difspec}). 
\section{Generalized noise channels in phase space} 
		\label{sec:channels}
In this section
we obtain many-qubit generalizations of the depolarizing and phase damping
channels  and write their superoperators in terms of  translations in phase
space.  With a slightly different approach these generalizations have been
studied in \cite{carlo}.

In \sect{sec:chordrep} we proposed a noise superoperator which is diagonal in
the chord representation and whose Kraus operators are proportional to
the translation operators. 
The main feature of this type of noise
is the (diagonal) supermatrix of coeficients $C(\alpha)$ 
(or equivalently the spectrum which from \equa{eq:spectDe} is 
$\widetilde C(\beta)$, i.e. the DFT of $C(\alpha)$). 
We introduce a parameter $\epsilon$ which, if we think the noise is due to a
coupling to an environment, it allows to 
continuously change from no coupling ($\epsilon=0$) to full-strenght coupling
($\epsilon=1$). Thus \equa{eq:deps} can be split 
as a convex sum of two superoperators
\begin{equation}
	\label{eq:convex}
\s_\epsilon=(1-\epsilon)\hT_0\odot\hT_0+
\frac{\epsilon}{N}\sum_{\alpha=0}^{N^2-1}C(\alpha)
\hT^{ }_\alpha\odot\hT^{\dag}_\alpha.
\end{equation}
 Hence $\s_\epsilon$ is also diagonal in the chord representation. Following
\equa{eq:spectDe}, the spectrum of 
$\s_\epsilon$ is given by
\begin{equation}
	\label{eq:gensptr}
\Sigma(\beta)=(1-\epsilon)+\epsilon\widetilde{C}(\beta)
\end{equation}
(with $\beta\equiv(q,p)$ and $q,p=0,\cdots,N-1$).
This means that the spectrum
takes the constant value $(1-\epsilon)$ for all $\hT_\beta$ plus an additional
$\epsilon\widetilde{C}(\beta)$. Moreover 
$\widetilde{C}(0)=1$ for $\s_\epsilon$ to be unital.

\subsection{Depolarizing Channel}
	\label{sec:GDC}
The depolarizing channel for a single qubit, leaves it unchanged with probability
$(1-\epsilon)$ and {\em depolarizes\/} it, which means that it leaves it in a
completely mixed state, with probability $\epsilon$.
The Kraus operators for one-qubit depolarizing channel
\cite{preskill,chuang} are
\begin{equation}
	\label{eq:1qbitKrausDC}
    \begin{array}{ll}
       \hat{M}_{0}&=\sqrt{(1-\epsilon)}\hat{I}  ,\\
       \hat{M}_{1}&=\sqrt{\frac{\epsilon}{3}}\hat{\sigma}_{1}=\sqrt{\frac{2\epsilon}{2(2^{2}-1)}}\hat{\sigma}_{1} ,\\
       \hat{M}_{2}&=\sqrt{\frac{\epsilon}{3}}\hat{\sigma}_{2}=\sqrt{\frac{2\epsilon}{2(2^{2}-1)}}\hat{\sigma}_{2} ,\\
       \hat{M}_{3}&=\sqrt{\frac{\epsilon}{3}}\hat{\sigma}_{3}=\sqrt{\frac{2\epsilon}{2(2^{2}-1)}}\hat{\sigma}_{3} .
    \end{array}
\end{equation}
If not stated explicitly matrices are written in the computational basis
$\{\ket{0},\ket{1}\}$.
Notice that each $\hM_{\mu}$  is a constant multiplied by
the corresponding Pauli matrix
$\hat\sigma_\mu$ (with $\hat\sigma_0\equiv\II$), 
which are the generators of the $SU(2)$
group and constitute an orthonormal basis of $\hil_{2^2}$. 
\begin{figure}[ht!]
\begin{center}
\scalebox{0.7}{
\includegraphics*[57,73][399,242]{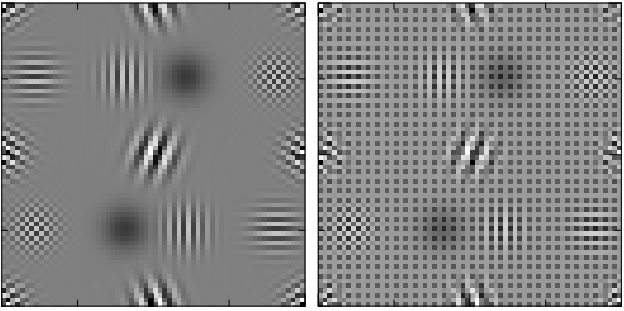}}
\caption{Wigner function representation of the 
action after one step of the depolarizing channel 
$\SOp{S}^{DC}$, with 
$N=32$ and  $\epsilon=0.9$. The left panel shows as initial state a cat state formed by the 
superposition of two coherent states centered at $(q,p)=(0.4,0.25)$ and $(0.6,0.75)$. 
Long wavelength interference on the center
corresponds to pure quantum interference while the othe short wavelength fringes
correspond to interference with images generated by the periodicity of the torus.   
The uniform averaging over all phase space is clearly seen on the right panel. 
\label{fig:dpc}}
\end{center}
\end{figure}
For $K$ qubits a straightforward generalization of \equa{eq:1qbitKrausDC}
can be constructed with the Kraus operators given by
\begin{equation}
    \begin{array}{ll}
       \hat{M}_{0}=\sqrt{(1-\epsilon)}\hat{I}  ,\\
       \hat{M}_{\mu}=\sqrt{\frac{N\epsilon}{2(N^{2}-1)}}\hat{\gamma}_{\mu} ,\\      
    \end{array}
\end{equation}
where $\gamma_\mu$ is the $\mu$th element of the 
set of $N^2-1$ generators of $SU(N)$. 
They are hermitian operators which satisfy
\begin{subequations}
   \begin{eqnarray}
      \tr\{\hg_\mu\}&=&0\\
      \tr\{\hg_\mu\hg_\nu\}&=&2\delta_{\mu\nu}.%
   \end{eqnarray}
\end{subequations}
The normalized operators 
\begin{subequations}
	\label{eq:Q}
   \begin{eqnarray}
      \hQ_0&=&\frac{1}{\sqrt{N}}\II \label{eq:Qa}\\
      \hQ_\mu&=&\hQ_\mu^\dag=\frac{1}{\sqrt{2}}\gamma_\mu
     \label{eq:Qb}
  \end{eqnarray}
\end{subequations}
form a complete basis in $\hilNN$.
We propose for the {\em generalized depolarizing channel\/} 
the following expresion
\begin{eqnarray}
	\label{eq:DC}
\s^{\text{DC}}_\epsilon&=&(1-\epsilon)\hat{I}\odot\hat{I}+
   \frac{\epsilon}{N}\sum_{\mu=0}^{N^{2}-1}\hQ_{\mu}\odot\hQ_{\mu}
\nonumber \\   
   &\stackrel{\text{def}}{=}&(1-\epsilon)\II\odot\II+\epsilon\s_2
\end{eqnarray}
Using the fact that the set
$\{\hQ_\mu\}_{\mu\ne 0}$ spans the same subspace of traceles operators as
$\{\hT_\alpha/\sqrt{N}\}_{\alpha\ne 0}$, it can be seen that $\s_2$ in
\equa{eq:DC} is trace preserving.  

The computational
basis, which we arbitrarily took to be the position basis in discrete phase
space, defines a ``canonical'' orthonormal basis in Liouville space
$\hilNN$. The elements are the skew projectors (transition operators)
\begin{equation} 
\hat{P}_{ij}=\ket{i}\brac{j}, 
\end{equation} 
with
$i,j=0,\cdots,N-1$. Using \equa{eq:tqp} it is easy to see that 
\begin{eqnarray} 
\label{eq:Pij} \hat{P}_{ij}&\equiv&\frac{1}{N}\sum_{q,p=0}^{N-1}
\tr\left[\hat{T}^{\dagger}_{(q,p)}\hat{P}_{ij}\right]
\hat{T}_{(q,p)}\nonumber \\
&=&\frac{1}{N}\sum_{p=0}^{N-1}e^{-i\frac{\pi}{N}p(i+j)}\hat{T}_{(i-j,p)} .
\end{eqnarray}
where
$1/\sqrt N$ is added for normalization. For clarity, we used the two indices
of $\Tqp$ explicitly. 

Now, the generators $\gamma_\mu$ can be written in the computational basis in terms of
skew projectors as (see, for example, \cite{Mahler})
\begin{eqnarray}
\gamma_\mu&\rightarrow\{\hU_{12},\hU_{13},\hU_{23},\cdots,\hV_{12},\hV_{13},
\hV_{23},\nonumber \\
&\cdots,\hW_1,\hW_2,\cdots,\hW_{N-1}\}
\end{eqnarray}
with
\begin{equation}
	\label{eq:gamma}
  \left\{ 
    \begin{array}{rl}
       \hU_{jk}=&\hat{P}_{jk}+\hat{P}_{kj}, \\ 
       \hV_{jk}=&i(\hat{P}_{jk}-\hat{P}_{kj}), \\  
       \hW_{l}=		
       &-\sqrt{\frac{2}{l(l+1)}}(\hat{P}_{11}+\hat{P}_{22}+\\
       &\cdots+\hat{P}_{ll}-l\hat{P}_{l+1,l+1})
    \end{array}
  \right.
  ,
\end{equation}
where $1\leqslant j<k\leqslant N$, $1\leqslant l\leqslant N-1$. 
Inserting
Eqs.~(\ref{eq:Pij}) and (\ref{eq:gamma}) into \equa{eq:DC}
it can be transformed into
\begin{equation} 
\s^{\text{DC}}_\epsilon=(1-\epsilon)\hat{T}^{ }_{0}\odot
  \hat{T}^{\dagger}_{0}+\frac{\epsilon}{N^{2}}\sum^{N^2-1}_{\alpha=0}
  \hat{T}^{ }_{\alpha}\odot\hat{T}^{\dagger}_{\alpha}.
\end{equation}
The physical interpretation is quite simple from a phase space point of view.
With probability $(1-\epsilon)$ it leaves the state unchanged while with uniform
probability $\epsilon/N^2$ it performs all the possible translations, thus averaging
over all phase space with equal weight (except at the origin). In FIG.~\ref{fig:dpc}
the action of $\s^{\text{DC}}$ on a superposition of two coherent states (represented
by the Wigner function) is shown, with $\epsilon=0.9$. 
The value of $\epsilon$ is taken purposely large to make the 
effects of the averaging become apparent. Eventually, further
averaging over the whole phase space
leads to a completely depolarized state, i.e. the uniform state $\II/N$. 
\begin{figure*}[ht!]
\begin{center}
\includegraphics*[60,605][529,752]{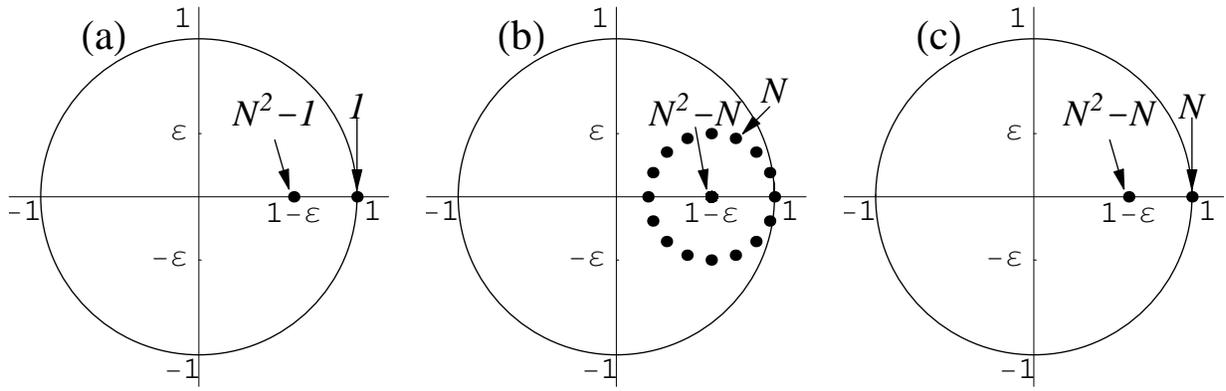}
\caption{Complex plane plot of (a) The spectrum of the generalized depolarizing channel. Notice the
$N^2-1$ degeneracy on $(1-\epsilon,0)$; (b) The spectrum for the Phase
damping channel on the line with 
$N=32$, $n_1=1,n_2=2,n_3=2$. There are $N^2-N$
eigenvalues on $(1-\epsilon,0)$ and $N$ (doubly degenerate) eigenvalues on the
circle of radius $\epsilon$ centered at $(1-\epsilon,0)$; and 
(c) The spectrum for the Phase
damping channel on the line with 
$N=32$, $n_1=1,n_2=0,n_3=2$. There are $N^2-N$
eigenvalues on $(1-\epsilon,0)$ and $N$ eigenvalues equal to 1.
\label{fig:auto}}
\end{center}
\end{figure*}

Equation (\ref{eq:gensptr}) gives the spectrum 
\begin{eqnarray}
\Sigma^{DC}(\beta)&=&(1-\epsilon)+\epsilon
\sum_{\alpha}\frac{1}{N^2} 
	e^{i(2\pi/N)\beta\wedge\alpha}\nonumber \\
	&=&(1-\epsilon)+\epsilon\delta_{\beta,0}.
\end{eqnarray}	
The spectrum takes the constant value $(1-\epsilon)$ for all $\Tqp$ except for 
$\hT_{(0,0)}$ where it is equal to 1 (FIG~\ref{fig:auto}, left).
The $(N^2-1)$-degeneracy makes the composition of this channel with any other
superoperator trivial. The effect is to contract all the spectrum, except for a
single 1, by a factor $(1-\epsilon)$. 

When $\epsilon\sim 1$ the spectrum lies within a small boundary of the origin leaving
1 as the unique non-zero eigenvalue. Physically this means that after one step all th
modes that are orthogonal to the identity decay. Thus the only preferred state in
this limit case is the uniform density $\II/N$.

\subsection{Phase-Damping}
The phase-damping channel is widely used in the contexts of quantum to classical
correspondence and quantum information because it provides a simple picture of how decoherence acts
picking out preferred sets of states \cite{preskill}.
The Kraus operators for the one-qubit phase damping channel are
\begin{equation}
    \begin{array}{rl}
       \hat{M}_{0}&=\sqrt{(1-\epsilon)}\hat{I} ,\\
       \hat{M}_{1}&=\sqrt{\epsilon}\ket{0}\brac{0} ,\\
       \hat{M}_{2}&=\sqrt{\epsilon}\ket{1}\brac{1} .
    \end{array}
\end{equation}
It is instructive to see what happens after repeated action of the phase-damping
channel. It is easy to check that after $n$ steps, the initial density matrix
$\hrho_0$ is 
\begin{equation}
(\s^{\text{PDC}}_\epsilon)^n(\hrho_0)=\left(
\begin{array}{cc}
\rho_{00}&(1-\epsilon)^n\rho_{01}\\
(1-\epsilon)^n\rho_{10}&\rho_{11}
\end{array}
\right).
\end{equation}
So if $\epsilon$ represents a decay rate, exponential decay of the non-diagonal terms
occurs leaving the state in a completely mixed state. Therefore decoherence 
picks out the computational states as preferred basis. Unlike the depolarizing channel, phase
damping only produces loss of coherence.

We propose a generalized $\s^{PDC}_\epsilon$ for $K$ qubits in terms of 
The skew projectors $\hat P_{ij}$ 
\begin{equation}
\s^{\text{PDC}}_\epsilon=(1-\epsilon)\II\odot\II+\epsilon\sum_{i,j=0}^{N-1}
C_{ij}\hat{P}^{ }_{ij}\odot\hat{P}^{\dag}_{ij}.
\end{equation}
To identify $\sqrt{\epsilon C_{ij}}\hat P_{ij}$ with Kraus operators
then the coefficients $C_{ij}$ must be real and positive.
Moreover, in order to be trace preserving the identity $\sum_{ij}C_{ij}=1$ 
must hold. Using the same arguments of Sec.~\ref{sec:GDC} an expresion 
in terms of translation operators 
can be obtained. For example, if $C_{ij}=\delta_{ij}$ we get
\begin{equation}
\s^{PDC}_\epsilon=(1-\epsilon)\II\odot\II+\frac{\epsilon}{N}
\sum_{p=0}^{N-1}\hT_{(0,p)}^{ }\odot\hT^{\dag}_{(0,p)}.
\end{equation}
\begin{figure}[ht!]
\begin{center}
\scalebox{0.7}{
\includegraphics*[57,73][399,242]{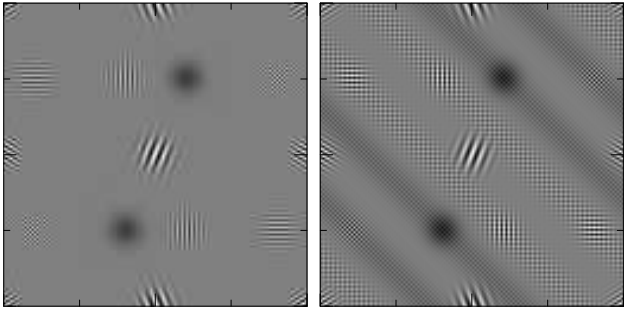}}
\caption{Wigner function representation of the 
action after one step of the phase damping on the line 
$\s_{\text{1,-1,0}}$, with 
$N=64$ and $\epsilon=0.85$. The initial state is the same as in 
FIG~\ref{fig:dpc}. The averaging over the line $p=-q$ (mod 64)
can be appreciated on the right panel. (Notice that the short
wavelength interferences are due to the torus periodicity.) \label{fig:pdc}}
\end{center}
\end{figure} 

The physical interpretation follows intuitively. With probability $(1-\epsilon)$ it
leaves the state unchanged, while it averages in a preferred direction (in this
case the vertical lines $q=\text{{\it const.}}$, $p\in[0,N-1]$). The channel picks
out as pointer basis the position projectors $\hat P_{ii}$. This is not a surprise
since we (arbitrarily) took as computational basis, the position eigenstates
$\ket{q}$. Had we chosen the momentum basis instead, then the averaging
over lines $p=const$ would lead to momentum as pointer basis.
Following this train of thought we
show that different pointer bases can be selected by ``diffusing'' along any line in
the phase space grid.
We define a line on the $N\times N$ grid $G_N$ of points $(q,p)$ in 
phase space as
\begin{equation}
L_{n_{1}n_{2}n_{3}}=\{(q,p)\in
G_{N};\,n_{1}p=n_{2}q+n_{3},\,n_{1},n_{2},n_{3}\in \Bbb{N}\}
\end{equation}
The generalized phase damping channel on the line 
$L_{n_{1}n_{2}n_{3}}$ is given by
\begin{eqnarray}
	\label{eq:PDCline}
  \s^{\text{PDC}}_{\text{$n_{1}n_{2}n_{3}$}}&=&
  (1-\epsilon)\hT_{(0,0)}\odot\hT^{\dagger}_{(0,0)}
  \nonumber \\
  & &
  +\frac{\epsilon}{R}\sum_{(q,p)
  \in L_{n_{1}n_{2}n_{3}}}\hT_{(q,p)}\odot\hT^{\dagger}_{(q,p)}\nonumber
  \\
  &=&(1-\epsilon)\II\odot\II+\epsilon\SOp{D}_{n_1n_2n_3},
\end{eqnarray}
where $\SOp{D}_{n_1n_2n_3}$ 
is a superoperator like (\ref{eq:deps}).
The integer $R$ is the number of points contained in the line. 
If either $n_1$ or  $n_2$ is odd then $R\equiv N$. 
On the other hand, it is an integer  
power of 2 times $N$ if both $n_1$ and $n_2$ are even \cite{Miquel}.
 
The eigenvalues corresponding to each translation $\Tqp$ follow from the
analysis in Sects.~\ref{sec:chordrep} and \ref{sec:channels}
\begin{equation}
	\label{eq:spPDC1}
\Sigma^{\text{PDC}}(q,p)=1-\epsilon(1-e^{-i(2\pi/N)qn_3/n_1}\delta_{n_2 q,n_1 p})
\end{equation}
for $n_1\ne 0$, while if $n_1=0$ and $n_2\ne 0$ it is
\begin{equation}
	\label{eq:spPDC2}
\Sigma^{\text{PDC}}(q,p)=1-\epsilon(1-e^{i(2\pi/N)p n_3/n_2}\delta_{q,0}).
\end{equation}
%
This spectrum is displayed in FIG.~\ref{fig:auto} (b) and (c).

In 
FIG.~\ref{fig:pdc} the effect of $\s^{\text{PDC}}_{\text{$n_1n_2n_3$}}$ 
acting on a superposition 
of coherent states can be seen, for the case $n_1=1,n_2=-1$ and $n_3=0$.
The right panel shows how the line ($p=-q$) 
that the channel picks, and averages over, appears. The only difference with a
line with non-zero $n_3$ would be a vertical translation 
of the line represented. Moreover, there is a  
progresive erasure of quantum interferences 
between the two coherent states. 

The action of this type of noise channel can be best understood if we delve 
deeper into the action of $\SOp{D}_{n_1n_2n_3}$. Using \equa{eq:Tdecomp} it can
be written as a composition of two superoperators
\begin{equation}
\SOp{D}_{n_1n_2n_3}=
\left\{
\begin{array}{ll}
\SOp{D}_{n_1n_20}\circ\SOp{V}^{n_3/n_1},&\ n_1\ne 0\\
 \SOp{D}_{0n_20}\circ\SOp{U}^{-n_3/n_2},&\ n_1=0
\end{array}
\right.
.
\end{equation}
where the divisions in the exponents should be understood as the product 
with the inverse of an integer, modulo $N$ (whenever such inverse exsists).
Therefore
$\SOp{V}^{n_3/n_1}=\hV^{n_3/n_1}\odot(\hV^\dag)^{n_3/n_1}$ is a unitary
translation of $n_3/n_1$ in momentum and 
$\SOp{U}^{-n_3/n_2}=\hU^{-n_3/n_2}\odot(\hU^\dag)^{-n_3/n_2}$ is a unitary translation of 
$-n_3/n_2$ in position.
In other words, first apply a translation in position or momentum and then
average on the line of slope $n_2/n_1$ that contains the origin. 

This superoperator has interesting features 
when $\epsilon\sim 1$. 
Equations (\ref{eq:spPDC1}) and (\ref{eq:spPDC2}) 
show that there is an $N^2-R$ dimensional null subspace. The remaining non-zero eigenvalues are on the unit 
circle and, depending on the parity of $n_1, n_2$ and $n_3$, may exhibit double 
(or some integer power of 2) degeneracy. 
The effect is that the averaging superoperator $\SOp{D}_{n_1n_20}$  
truncates 
the unitray superoperator $\SOp{V}^{n_3/n_1}$ to a smaller  
submatrix which is still unitary (the same applies
to $\SOp{D}_{0n_20}$ composed with $\SOp{U}^{-n_3/n_2}$).

When this noise channel acts on a unitary map (with $\epsilon\sim 1$)
the full propagator is truncated to an $R$-dimensional subspace spanned
by translations on the chosen line.
\section{Diffusive Noise}
	\label{sec:difnoise}
Although the effect of the previous two types of noise is clear in the context of 
many-qubit quantum algorithms, it is not intuitive to picture them as 
diffusion in phase space, because they average over large distances 
(namely the whole space). 
On the other hand, although a purely diffusive noise is difficult
to ``see'' acting on a system of qubits, 
it has a natural interpretation in phase space.
However, in spite of this differences, they can be treated as different types (depending
on a function $C(\alpha)$) of the
same class of noise given in \equa{eq:convex}.
\begin{figure}[!htb]
\begin{center}
\scalebox{0.5}{
\includegraphics*[143,207][463,521]{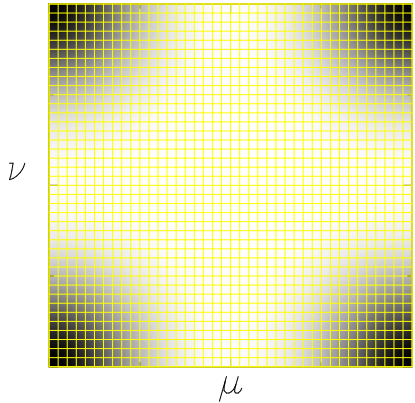}}
\caption{Schematic plot of $\cmn$.  The torus  phase space is equivalent to a
square with periodic boundary conditions. The points on the grid represent
the states.
The size of the dark region 
represents  the effective rank of the truncated
$\EuFrak{U}_{(p',q')(p,q)}$ (\equa{eq:trunc}).
Numerically we set the white part equal to 0.  \label{fig:shadow}}
\end{center}
\end{figure}
Following recent works \cite{garma,garma2,nonn} 
we define as diffusive noise, 
\begin{equation}
\De=\sum_{\alpha}\ca\hT_{\alpha}\odot\hT^{\dag}_{\alpha}
\end{equation}
which is a  
particular case of  \equa{eq:convex} with $\epsilon=1$ and  
where $C_\sigma(\alpha)$ is a
periodic function narrowly peaked around $\alpha=0$ 
and of approximate width $\sigma$. 
For practical purposes we take $\ca$ to be a periodized 
Gaussian of 
half-width $\sigma$ 
and thus $\cl$ is also a Gaussian of half-width $1/(2 \pi \sigma N)$ 
(FIG.~\ref{fig:shadow}, where $\lambda\equiv (\mu,\nu)$).
\begin{figure*}[htb!]
\begin{center}
\begin{turn}{-90}
\scalebox{0.85}{
\includegraphics*[202,116][412,674]{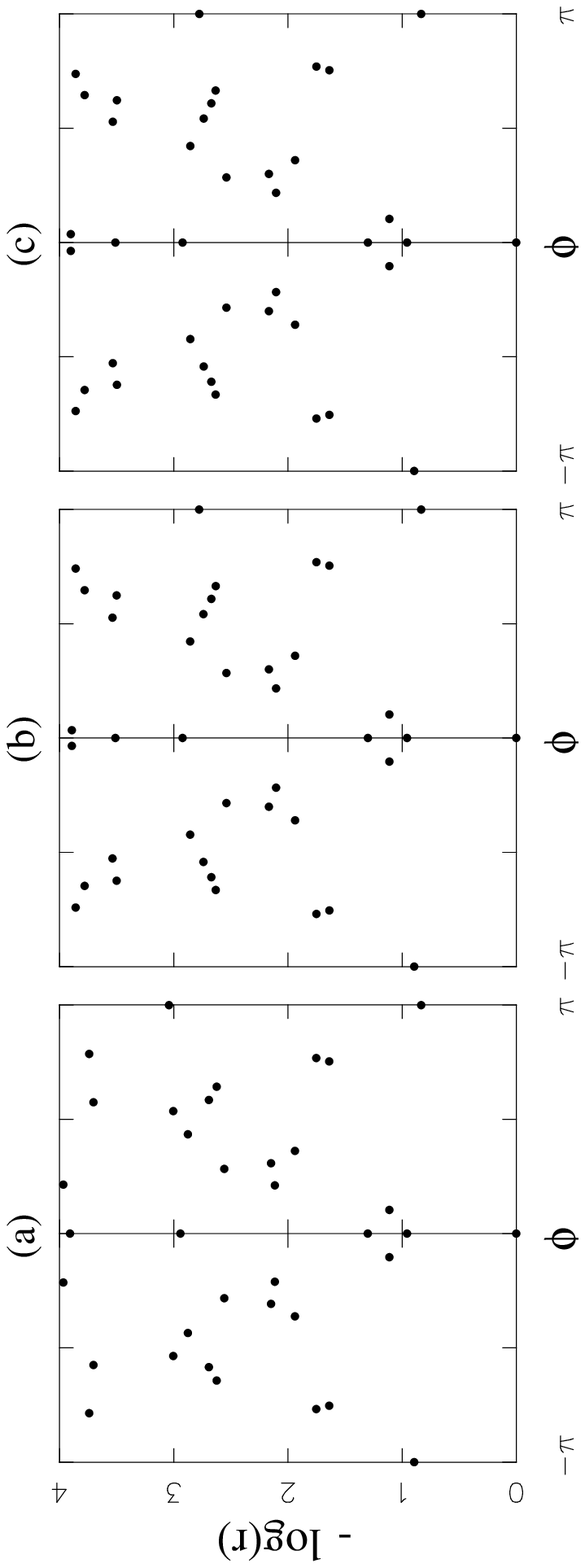}}
\end{turn}
\caption{Leading spectrum of $\doll=\De\SOp{U}$, where $\De$ 
is the difussive noise of
\sect{sec:difnoise} (with $\sigma=0.063$) 
and $\SOp{U}$ is the unitary superoperator for 
the quantum perturbed
Arnold cat used in \cite{garma,garma2} (with  
$N=100$, $k=0.02$). The dimension of the propagator  after the truncation is 
$dim=4(a/(2\pi \sigma))^2$. (a) $a=2$, $dim=4(2/2 \pi\sigma)^2\approx 100$; (b) $a=2.8$ 
and $dim=4(2.8/2 \pi\sigma)^2\approx 196$; (c) $a=4.8$ and $dim=4(4.8/2 \pi\sigma)^2\approx 576$.
(Note that $N^2\times N^2= 10000$.) 
If $\lambda_i$ is the $i$-th eigenvalue, then
$\log\lambda_i=\log(r_i)+ i \phi_i$ (where $r_i=|\lambda_i|$)
and the coordinates in the plots are
$(\phi,-\log(r))$.\label{fig:spectr}}
\end{center}
\end{figure*}
This implements an incoherent sum of all the possible
translations in phase space weighed by $C_\sigma(\alpha)$. 
To avoid a net drift $C_\sigma(\alpha)$ should satisfy 
\begin{equation}
C_\sigma(\alpha)=C_\sigma(-\alpha)
\end{equation}
which makes $\De$ a Hermitian operator.

In the context of quantum optics a similar type of noise for continuous phase space
variables has been called {\em Gaussian noise\/} \cite{hall}.
\subsection{Spectral properties of the noisy propagator}
	\label{sec:difspec}
In the examples of \sect{sec:channels} the effect of noise acting on the
unitary propagation was very simple. On the other hand, in the case of
diffusive noise the numerical calculation is more involved and its nature
depends strongly on the properties of the unitary map. 
In the context of quantum classical
transition the spectrum of the noisy propagator, for chaotic systems,  
has gained interest because in the limits of $\hbar\to 0$ and no noise it can
be related to purely classical properties like the Ruelle-Pollicott (RP)
resonances \cite{blank, nonn}. The RP resonances determine the long time
decay of classical quantities \cite{ruel} and as was recently shown
\cite{garma2} they determine the long time aproach to equilibrium of such
quantities as the purity and fidelity for
classicaly chaotic quantum systems with noise. 
Experimentally the appearance of RP resonances in quantum systems has been
shown, for example, in \cite{sridhar}.
Clearly the part of the
spectrum that can be related to long time decays is the 
largest in modulus (and smaller than 1).

The chord representation is specially suited for these calculations
because since $\ca$ is a Gaussian of  half-width $\sigma$ then the spectrum 
$\cl$ is also a Gaussian of complementary half-width $1/(2 \pi \sigma)$ 
which means that for some
ranges of $\sigma N$ there is a large part of the spectrum of $\De$
very close to zero.

The large ($N^2\times N^2$) supermatrix of the unitary step in the
chord representation is 
\begin{equation}
\EuFrak{U}_{\alpha'\alpha}=\tr\left[ 
 \hT_{\alpha'}\hU\hT_\alpha\hUd\right]
 \end{equation}
Since the noise superoperator is diagonal in this representation its
application is particularly simple and is given by
 \begin{equation}
    \label{eq:trunc}
\EuFrak{L}_{\lambda'\lambda}=\widetilde{C}_{\sigma}(\lambda')
\EuFrak{U}_{\lambda'\lambda}
\end{equation}
Thus for $\lambda'\gg 1/(\sigma N)$ the matrix elements of \equa{eq:trunc}
are negligibly small. Therefore $\cl$ tends to kill 
the long chord components of the unitary step. 
A schematic representation of $\cl$ is found in FIG.~\ref{fig:shadow}. The grid
represents points in phase space.  
The whiter region
gives the idea of which matrix elements can be
neglected. 
The number of points inside the darker region represents the  rank 
of the noisy superoperator reduced by the diffusion.

In the numerical treatment
we pick a threshold of values for $\cl$. To do so we determine a range which is a
coeficient $a$  times the width of $\cl$. Outside this range the quai-null spectrum is set
exactly equal to zero. The resulting matrix is of a size of order $4 (a/(2 \pi \sigma
N))^2$ 
(see FIG.~\ref{fig:shadow}). This procedure provides, 
up to a reasonable
accuracy, the largest part of the spectrum of $\doll$.  
The
calculation becomes more difficult as $\sigma \to 0$ and to
reobtain the unitary spectrum the full dimension $N^2\times N^2$
is needed. 
For chaotic maps the spectrum is composed of the eigenvalue 1, corresponding 
to $\hrho_0=\II/N$, a small set of eigenvalues with modulus smaller than one, 
and many eigenvalues close to zero. The action of
the noise is to set the latter ones exactly equal to zero. 

In FIG~\ref{fig:spectr} we see computation of the 
spectrum for three different sizes of
truncation for a quantum map with diffusion.
The leading part of the spectrum is unchanged while ther appear
slight in the less significant part.
The stability of the spectrum
up to orders of $10^{-5}$ can be clearly observed. 

Many different schemes have been used to calculate the leading spectrum of
the noisy porpagator\cite{garma,haak, agam, nonn,florido}. 
The advantage over these methods is
that although they use smaller matrices, the number of eigenvalues obtained is limited by accuracy
and diffucult error estimation. 

In \cite{spina} there is an analogous analysis for diffusive noise on a
spherical phase space.
\section{Conclusions}
We provided a generalization of two well known noise channels in the context of
phase space representations of quantum mechanics. Specifically we showed how the
depolarizing and phase damping channels have superoperator expressions that are very
simple to study using the chord representation. Moreover their spectral features,
which can be crucial when composed with a map or an arbitrary algorithm  are well
determined. Some useful properties of these generalized noise channels composed with
quantum maps and algorithms are currently being studied \cite{colo}.

The noises where formulated as a one parameter ($\epsilon$) family and
their spectral properties where studied. In particular 
when the spectrum of the noise has a large null eigensubspace the convolution
with a unitray map results in a truncated matrix whose effective 
rank depends on the rank of the noise.
Thus reducing the computational requirements for systems with a large
Hilbert space.
This was shown to be useful when computing the leading spectrum of coarse
grained propagators corresponding to classsically chaotic quantum maps. The
coarse graining was obtained by means of a diffusive noise and we showed 
that the leading spectrum is independent of the truncation.

 \begin{acknowledgments}
This work was financially  supported by CONICET and ANPCyT. 
M.L.A. thanks the ``Pedro F. Mosoteguy''
Foundation for partial support.
\end{acknowledgments}

%

\begin{thebibliography}{99}
\bibitem{davi} 
L. Davidovich, in {\em Latin American School of Physics\/} XXXI ELAF, edited
by Shahen Hacyan, Roc\'{\i}o J\'auregui, and Ram\'on L\'opez-Pe\~na, AIP Conf.
Proc. 464 (AIP, Woodbury, NY,1999); Proceedings of the first PASI Conference on
Chaos, Decoherence and Entanglement 2000, available at
\url{http://kaiken.df.uba.ar.}
\bibitem{zurek1}
See, for example, W. H. Zurek, Phys. Today {\bf 44} (10), 36 (1991); J. P. Paz, S.
Habib, and W. H. Zurek, Phys. Rev. D {\bf 47}, 488 (1992).
\bibitem{paz1}
J. P. Paz and W. H. Zurek, in {\em Coherent Matter Waves, Proceedings of the Les
Houches Session LXXII\/}, edited by R. Kaiser, C. Westbrook, and F. David (Springer
Verlag, Berlisn, 2001).
\bibitem{Miquel} C. Miquel, J. P. Paz and M. Saraceno,
Phys. Rev. A, {\bf 65}, 2309 (2002).
\bibitem{zurek_nature}
W. H. Zurek, Nature {\bf 412}, 712 (2001).
\bibitem{ozorivas} A.M. Ozorio de Almeida Phys. Rep. {\bf 295} 266
(1998); A. Rivas and  A. M. Ozorio de
Almeida, Ann. Phys. (N.Y.){\bf 276}, 223 (1999).
\bibitem{preskill}
J. Preskill. ``1998 Lecture Notes for Physics 229: Quantum Information and
Computation'' available
at \url{http://www.theory.caltech.edu/people/preskill}.
\bibitem{chuang}
I. Chuang and M. Nielsen. {\em Quantum Information and Computation\/}
(Cambridge University Press, Cambridge, UK, 2001).
\bibitem{garma}
I. Garc\'\i a-Mata, M. Saraceno, and M. E. Spina,
Phys. Rev. Lett. {\bf 91}, 064101 (2003).
\bibitem{garma2}
I. Garc\'\i a-Mata and M. Saraceno, {\bf 69}, 056211 (2004).
\bibitem{L} 
G. Lindblad, Commun. Math Phys, {\bf 48}, 119 (1976).
\bibitem{qnoise}
G. W. Gardiner and P. Zoller, ``Quantum Noise: A
Handbook of Markovian and Non-Markovian Quantum Stochastic Methods with
Applications to Quantum Optics'', Springer-Verlag, Berlin-Heidelberg 2000;
B. K\"ummerer. {\em Quantum Markov Processes\/} in ``Coherent
Evolution in Noisy Environments'', A. Buchleitner and K.
Hornberger (Eds.), Springer-Verlag, Berlin-Heidelberg (2002).
\bibitem{GKS} V. Gorini, A. Kossakowski, and E. C. G. Sudarshan, J. Math. Phys. {\bf 17},
821 (1976).
\bibitem{Kraus} K. Kraus, {\it States, Effects and Operations}, Springer-Verlag,
Berlin, 1983.
\bibitem{havel}
See T. F. Havel, J. Math. Phys. \textbf{44}, 534 (2003), and references therein.
\bibitem{voros}
G. Palla, G. Vattay and Andre Voros, Phys. Rev. E {\bf 64}, 012104 (2001).
\bibitem{braun}
D. Braun, CHAOS, \textbf{9},730 (1999) and  D. Braun, Physica D,
\textbf{131}, 265 (1999); 
D. Braun, ``Dissipative Quantum Chaos and Decoherence'', Springer-Verlag,
Berlin-Heidelberg (2001).
\bibitem{nonn}
S. Nonnenmacher, Nonlinearity \textbf{16}, pp.1685-1713 (2003).
\bibitem{spina}
M. E. Spina and M. Saraceno, e-print nlin.CD/0406001.
\bibitem{nature}
C. Miquel, J.P. Paz, M. Saraceno, E. Knill, R. Laflamme, and C. Negrevergne, Nature
{\bf 418}, 59 (2002).
\bibitem{Paz}
J. P. Paz, A. J. Roncaglia, and M. Saraceno, Phys. Rev. A \textbf{69}, 032312
(2004).
\bibitem{schwinger}
J. Schwinger, {\em Proc. Nat. Acad. Sci.\/} \textbf{46}, 570 (1960).
\bibitem{hannay}
J. H. Hannay and M. V. Berry, Physica 1D,267-290 (1980).
\bibitem{knill}
E. Knill, Los Alamos National Laboratory Report LAUR-96-2717 and 
Los Alamos National Laboratory Report LAUR-96-2807
\bibitem{carlo}
G. G. Carlo, G. Benenti, G. Casati, and C. Mej\'\i a-Monasterio
Phys. Rev. A {\bf 69}, 062317 (2004).
\bibitem{Mahler}
\selectlanguage{german}
G. Mahler and V. A. Weberru"sß, 
\selectlanguage{english}
{\em Quantum Networks: Dynamics of Open Nanostructures.} 
(Springer Verlag, Berlin, 1998).
\bibitem{hall}
M. J. Hall, Phys. Rev. A {\bf 50}, 3295 (1994).
\bibitem{blank}
M. Blank, G. Keller, and C. Liverani, Nonlinearity \textbf{15}, pp.1905-1973
(2002).
\bibitem{ruel}
D. Ruelle, Phys. Rev. Lett. \textbf{56}, 405 (1986);
D. Ruelle,  J. Stat. Phys \textbf{44}, 281 (1986).
\bibitem{sridhar}
K. Pance, W. T. Lu and S. Sridhar, Phys. Rev. Lett. \textbf{85}, 2737 (2000).
\bibitem{agam}
G. Blum and O. Agam, Phys. Rev. E \textbf{62}, 1977 (2000).
\bibitem{florido}
R. Florido, J. M. Mart\'{\i}n-Gonz\'alez and J. M. G\'omez Llorente, Phys. Rev.
E {\bf 66}, 046208 (2002).
\bibitem{haak}
C. Manderfeld, J. Weber, and F. Haake, J. Phys. A \textbf{34}, 9893 (2001).
\bibitem{colo}
M. L. Aolita and M. Saraceno, in preparation.
\end{thebibliography}
\end{document}